**Magnetic ordering with an XY-like anisotropy in the honeycomb lattice iridates $ZnIrO_3$ and $MgIrO_3$ synthesized via a metathesis reaction**

Yuya Haraguchi[1, 2, *], Chishiro Michioka[1], Hiroaki Ueda[1], Akira Matsuo[2], Koichi Kindo[2], Kazuyoshi Yoshimura[1, 3, 4, †]

[1]Department of Chemistry, Graduate School of Science, Kyoto University, Kyoto 606-8502, Japan
[2]The Institute for Solid State Physics, The University of Tokyo, Kashiwa, Chiba 277-8581, Japan
[3]Research Center for Low Temperature and Materials Sciences and [4]Institute of Liberal Arts and Science, Kyoto University, Kyoto 606-8501, Japan
*chiyuya@issp.u-tokyo.ac.jp
†kyhv@kuchem.kyoto-u.ac.jp

**abstract**

We have successfully synthesized the novel antiferromagnets with $Ir^{4+}$ honeycomb lattice $ZnIrO_3$ and $MgIrO_3$ and investigated their magnetic and thermodynamic properties. The two iridates are isomorphic but exhibit qualitatively different magnetic properties. $ZnIrO_3$ shows antiferromagnetic ordering below 46.6 K, whereas $MgIrO_3$ displays weak ferromagnetic behavior below 31.8 K owing to formation of a canted antiferromagnetic ordering. The measurement of magnetic susceptibility with using an oriented powder sample revealed the presence of an XY-like magnetic anisotropy and a tilting magnetic structure which is possibly stabilized by the Kitaev interaction. Moreover, magnetization curves of $MgIrO_3$ and $ZnIrO_3$ up to 60T show different behaviors, demonstrating that each magnetic ground state is different with each other. We discuss the difference in the ground state between $MgIrO_3$ and $ZnIrO_3$ from the viewpoint a magnetic model consisting of the Kitaev and Dzyaloshinskii-Moriya interactions with the spin frustration effect on the honeycomb lattice.

**Introduction**

Recently, physical properties driven by the spin-orbit coupling (SOC) have been attracted much attention from theorists and experimentalists [1-10]. In general, the electronic state of a $3d$ transition metal compound is hardly affected by SOC due to a strong effect of crystal field. On the other hand, in the case of $5d$ transition metal compounds, strong SOC would lead a characteristic electronic state. In this situation, SOC makes a vast change on the electron state. Especially in a low spin $d^5$ configuration like as $Ir^{4+}$, a $5d$ $t_{2g}$ band is split into half-filled $J_{eff} = 1/2$ and fully-filled $J_{eff} = 3/2$ bands by SOC. The half-filled $J_{eff} = 1/2$ state forms a narrow band, which results in that even small correlation makes a Mott gap [8]. This characteristic $J_{eff} = 1/2$ Mott state is firstly observed in the layered perovskite iridate $Sr_2IrO_4$ [8] and opened a new field of research in a quantum physics.

In the $J_{eff} = 1/2$ Mott state, a Kitaev interaction, recognized as an unconventional bond-directional ferromagnetic interaction, is theoretically predicted to be effective [9]. Indeed, a weak ferromagnetic transition in $CaIrO_3$ is explained by the Kitaev interaction

[11,12]. In the situation of presence of the Kitaev interaction on the honeycomb lattice, the ground state is exactly solved to be a quantum spin liquid [2]. In realistic compounds $Li_2IrO_3$, $Na_2IrO_3$, and $RuCl_3$, it is though that the Kitaev interaction in the honeycomb lattice is realized in a certain strength, and many intensive theoretical and experimental investigations have been performed [13-23]. Although no quantum spin liquid ground state is observed in these compounds, unconventional antiferromagnetic orderings were observed [24-29], of which possible origin has been proposed to be a coexistence of a Kitaev ferromagnetic interaction with Heisenberg antiferromagnetic interactions. In the case of $RuCl_3$, a fractional Kitaev spin liquid behavior is observed just above $T_N$ and under high magnetic fields [30, 31]. Most recently, it is reported that $H_3LiIr_2O_6$ prepared by a soft chemistry technique replacing interlayer lithium ions of $Li_2IrO_3$ with protons shows a spin liquid behavior down to 50 mK [32,33]. However, what the Kitaev interaction brings is not completely understood. Thus, in this research field, it is strongly required to find new iridium oxides with a honeycomb lattice, which would lead to a discovering exotic physics driven by the strong SOC.

In this paper, we report on synthesis of new iridates $ZnIrO_3$ and $MgIrO_3$ to be a model system of $J_{eff} = 1/2$ honeycomb lattice antiferromagnets via a low-temperature metathetical decomposition of $Li_2IrO_3$ and $ACl_2$ ($A$ = Mg, Zn) as well as the results of an investigation of their physical properties. Both compounds belong to a family of the ilmenite-type compounds with a regular honeycomb magnetic lattice formed by $Ir^{4+}$ ions. It is found that both of them exhibit antiferromagnetic order and in the case of $MgIrO_3$, the magnetic order occurs accompanied by a weak ferromagnetic moment. From the magnetic measurement with using the magnetic-field oriented samples, it was revealed that $ZnIrO_3$ and $MgIrO_3$ have XY-like magnetic anisotropy with an unconventional magnetic ground state, in which the ordered moment is along neither in the *c*-axis nor in the *ab*-plane. Based on the detailed magnetic measurements, we discuss the mechanism of the observed magnetic behaviors from a viewpoint of the Kitaev and Dzyaloshinskii-Moriya interactions with the spin frustration on the honeycomb lattice.

**Experiments**

A precursor material $Li_2IrO_3$ was prepared by the conventional solid-state reaction method in air. The obtained precursors were ground well with an excess of $ACl_2$ ($A$ = Mg, Zn) in an $N_2$-filled glovebox, sealed in an evacuated silica tube, and reacted at 400°C for 100 h. This metathetical reaction is expressed as,

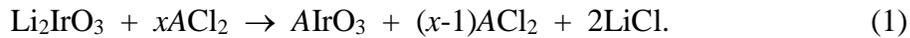

$$Li_2IrO_3 + xACl_2 \rightarrow AIrO_3 + (x-1)ACl_2 + 2LiCl. \qquad (1)$$

The residual $ACl_2$ and the byproduct LiCl were removed by washing with distilled water. These samples were characterized by the powder X-ray diffraction (XRD) on a diffractometer (RINT-2000; Rigaku) with Cu $K\alpha$ radiation. The cell parameters and the crystal structures were refined by the Rietveld method using RIETAN-FP v2.16 [34]. Magnetically oriented samples were prepared as following; the powder samples were embedded in an epoxy resin adhesive glue and submitted to a rotating magnetic field of $H$ =

5 T at room temperature until adhesive curing. In order to check how the oriented samples are well oriented, the XRD pattern of the oriented powder was measured in the condition of the scattering vector parallel to the axis of rotating magnetic field. As shown in Fig. 2(b), almost only 00$l$ diffractions can be obtained in XRD patterns of the oriented samples indicating a good orientation in the 00$l$ direction. The temperature dependence of the magnetization was measured under several magnetic fields up to 7 T by using a magnetic property measurement system (MPMS; Quantum Design) equipped at the LTM Research Center, Kyoto University. The temperature dependence of the specific heat was measured by using a conventional relaxation method with a physical property measurement system (PPMS; Quantum Design). Magnetization curves up to 60 T were measured using an induction method with a multilayer pulsed magnet at the International Mega Gauss Science Laboratory of the Institute for Solid State Physics at the University of Tokyo.

**Results and Discussion**

X-ray powder diffraction patterns of $ZnIrO_3$ and $MgIrO_3$ are shown in Fig. 2(c). All the indexed peaks without unknown impurity peaks for both compounds can be characterized using the ilmenite-type structure. The peaks are considerably broad, indicating a small particle size on the order of ~0.1 micron estimated using the Scherrer equation [35] The chemical composition both of compounds examined by an energy dispersive x-ray spectrometry was $A$/Ir ≈ 1, indicating a good stoichiometry. The structures of $ZnIrO_3$ and $MgIrO_3$ were refined by using the Rietveld method as described in the experimental section. Details of the refinement parameters are given in Table I. All nearest neighbor bond lengths between Ir ions are equivalent consistent with their space group of $R$-3. Thus, in this system, Ir ions form a perfect honeycomb lattice. The bond valence sum calculation [36] for Ir ions gave +3.80 and +3.92, respectively for $ZnIrO_3$ and $MgIrO_3$. These values reasonably agree with the expected valence of +4 for Ir ions. Bond angles of Ir-O-Ir, which is an exchange path among nearest neighbor Ir ions, are 94.0(3)° for $MgIrO_3$ and 95.7(1)° for $ZnIrO_3$. The vicinity to an ideal 90° angle of Ir-O-Ir possibly gives rise to effective Kitaev coupling [9].

The temperature dependence of the magnetic susceptibility $\chi$ for $ZnIrO_3$ and $MgIrO_3$ are shown in Fig. 3(a). As shown in the inset, there is a linear relationship in $1/\chi$ versus $T$ at high temperatures. The obtained parameters by the Curie-Weiss fitting with using the formula,

$$\chi = C/(T-\theta_W) = p_{eff}^2/8(T-\theta_W), \qquad (2)$$

where $C$ is a Curie constant, $p_{eff}$ is an effective paramagnetic Bohr magneton, and $\theta_W$ is the Weiss temperature, are listed in the TABLE II. The estimated $p_{eff}$ values are well coincide with the value of √3 ~ 1.73 expected for $J_{eff} = 1/2$ of $Ir^{4+}$ ions which is consistent with the fact that spins are well localized.

At low temperatures, the $\chi$ curves of both $ZnIrO_3$ and $MgIrO_3$ show anomalies, indicating magnetic orderings. The magnetic transition temperature $T_N$ are 46.6 and 31.8 K, respectively for $ZnIrO_3$ and $MgIrO_3$. Their behaviors of $\chi$ below $T_N$ have a marked difference. In the case of $ZnIrO_3$, the value of $T_N$ has approximately the same value of $\theta_W$, giving frustration index $f = |\theta_W|/T_N$ of approximately 1. In the lowest temperature, $\chi$ for $ZnIrO_3$ decreases to approximately two-thirds of that at the transition temperature, which is a typical behavior of a polycrystalline sample for three-dimensional antiferromagnetic ordering. On the other hand, the value of $T_N$ for MgIrO$_3$ are approximately half of $\theta_W$, giving $f$ of approximately 2. This fact suggests in MgIrO$_3$, the magnetic ordering is suppressed by the spin frustration. MgIrO$_3$ shows a steep increase in $\chi$ below $T_N$. Figure 3(b) shows the $M/H$ values under various magnetic fields plotted as a function of $T$. The value of $\chi$ under $\mu_0 H = 0.1$ T rapidly increases with decreasing temperature below $T_N$, followed by a thermal hysteresis between the zero-field-cooled and field-cooled data. With increasing the applied magnetic field, the increase of $M/H$ below $T_N$ is suppressed, which indicates a presence of ferromagnetic moment. In a temperature region near 0 K, $\chi$ of MgIrO$_3$ shows a Curie tail, which corresponds to 3% of Ir$^{4+}$ local moments ($S = 1/2$). The magnitude of this Curie tail is very small and sample dependent, which strongly suggests it is extrinsic. As shown in the inset of Fig. 4, an isothermal hysteresis is observed in the $M$-$H$ curve at 2 K. The spontaneous magnetic moment is about 0.0008 $\mu_B$ per Ir$^{4+}$ atoms. This very small moment is induced by the canted antiferromagnetic structure. Thus, the observed thermal hysteresis in $M/H$ should not to be due to a spin glass transition but related to the formation of ferromagnetic domains.

In the measurement of $\chi$ with using the oriented samples, the presence of magnetic anisotropy is revealed. Figure 4 shows the temperature dependence of the magnetic susceptibility $\chi$ in ZnIrO$_3$ and MgIrO$_3$ measured in the magnetic field parallel and perpendicular to the the $c$-axis. There are large magnetic anisotropies in $\chi$ in the whole temperature range. In ZnIrO$_3$ and MgIrO$_3$, the magnetic susceptibility in $H \perp c$ is larger than that in $H \parallel c$, suggesting an XY-like magnetic anisotropy. The highly anisotropic $g$ factors listed in TABLE II is derived from the low-spin state of Ir$^{4+}$ with the orbital degree of freedom in the trigonal crystal field. The magnetic anisotropy in ZnIrO$_3$ is larger than that in MgIrO$_3$, which is caused by a different trigonal distortion in IrO$_6$ octahedra. In a theoretical calculation, the anisotropic $g$ factor in a low spin $d^5$ ion with a trigonal distortion depends only on a trigonal splitting energy divided by the coefficient of spin-orbit interaction $\Delta_{tri}/\lambda$ [37]. Our best estimation from the anisotropic $g$ factor gives $\Delta_{tri}/\lambda=0.03$ for MgIrO$_3$ and $\Delta_{tri}/\lambda=0.14$ for ZnIrO$_3$. Generally, the value of $\Delta_{tri}/\lambda$ becomes larger with increasing a strength of distortion in a IrO$_6$ octahedron. Indeed, the octahedron in MgIrO$_3$ is more distorted than that in ZnIrO$_3$ as shown in the quadratic elongation of 1.0050 and 1.0207 for MgIrO$_3$ and ZnIrO$_3$, respectively, which calculated using the atomic displacement parameters. Such a situation is consistent with the difference of $\Delta_{tri}/\lambda$ and magnetic anisotropy.

Figure 5 shows the temperature dependence of the specific heat divided by temperature $C/T$ for ZnIrO$_3$ and MgIrO$_3$. To extract the magnetic contribution $C_M/T$, the lattice contributions have been estimated by fitting to the total $C/T$ data in the high temperature region with using the equation $C_{latt}/T =3R\{aC_D/T + (5-a)C_E/T\}$, where $R$ is the gas constant, $C_D$ and $C_E$ are the Debye- and Einstein-type heat capacities, and $a$ is the weight parameter. The best fit results are shown in the dashed lines with the parameters $a = 1.00(2)$, the Debye temperature $\Theta_D = 338(7)$ K, and the Einstein temperature $\Theta_E = 967(17)$ K for ZnIrO$_3$, and $a = 1.10(3)$, $\Theta_D = 461(9)$ K, and $\Theta_E = 1154(37)$ K for MgIrO$_3$. Each obtained $C_M/T$ grows below 100 K and shows a broad but distinct peak at $T_N$ each in ZnIrO$_3$ or MgIrO$_3$. The peak of $C_M/T$ indicates the entropy release associated with the magnetic long-range orderings. The small particle size probed by the XRD analysis may suppress divergence of the correlation length at the critical point, and this suppression would make the peak of $C_M/T$ at $T_N$ broadening. The magnetic entropy $S_M$ is calculated by integrating $C_M/T$. In the case of ZnIrO$_3$, $S_M$ reaches approximately 3.88 J mol$^{-1}$K$^{-1}$ at $T_N$. This value is 67% of the ideal total magnetic entropy of the $S = 1/2$ system. On the other hand, $S_M$ of MgIrO$_3$ reaches approximately 1.556 J mol$^{-1}$K$^{-1}$ at $T_N$, which value is 27% of the ideal one. In the case of MgIrO$_3$, a large part of the magnetic entropy would be released by the short-range magnetic correlation above $T_N$. This fact indicates that a spin correlation in MgIrO$_3$ is more frustrated than that in ZnIrO$_3$, which is consistent with values of $f = |\theta_W|/T_N$.

To clarify the difference between the magnetism in ZnIrO$_3$ and MgIrO$_3$, we have conducted a magnetization measurement up to 60 T at 4.2 K as shown in Fig. 5. We found field induced transitions in the magnetization processes both in ZnIrO$_3$ and MgIrO$_3$, whose behavior is different from each other. In the case of MgIrO$_3$, we observed a slight anomaly at $\mu_0 H_{SF}$ = 12 T in the differential susceptibility curve $dM/dH$, indicating a field-induced magnetic transition. In the case of ZnIrO$_3$, a rapid change of slope in $M$ at $\mu_0 H_a$ = 14.8 T is observed. The increase of $M$ above $\mu_0 H_a$ would correspond to a canted magnetic structure in which a canted angle becomes larger with increasing applied field. No jump of $M$ indicates that the transition is of a second order. The magnetization behaviors of MgIrO$_3$ and ZnIrO$_3$ are apparently different, which indicates that the magnetic ground states in ZnIrO$_3$ and MgIrO$_3$ are different.

Here, we discuss the magnetic structure on the bases of behavior of $\chi_c$ and $\chi_{ab}$ below $T_N$. In the case that the direction of all the moments are parallel to a certain axis, $\chi_{//}$ decreases and goes to zero below $T_N$, while $\chi_\perp$ is almost constant down to 0 K. In the present compounds, neither $\chi_c$ nor $\chi_{ab}$ shows a constant behavior below $T_N$, suggesting that the collinear ordered moment does not point the crystalline principal axis, or the magnetic structure is not a collinear one. In the present stage, it is not clear whether the magnetic structures are collinear or noncollinear. In the case of a similar honeycomb system Na$_2$IrO$_3$ and RuCl$_3$ with an XY anisotropy, similar behaviors of $\chi_c$ and $\chi_{ab}$ were explained by the collinear zig-zag magnetic structure with the spin-tilting out of the $ab$-plane [28,29]. The origin of spin-tilting in Na$_2$IrO$_3$ and RuCl$_3$ is theoretically predicted to be a compromise among Kitaev and Heisenberg interactions, in which the direction of spin tilting points along a $M$-$X$ direction [28,29]. Thus, in the present system, similar tilting magnetic structure could be realized.

It is reasonable to think that the spin tilting is stabilized by the Kitaev interaction. Next, we discuss the origin of weak ferromagnetic moment observed not in ZnIrO$_3$ but in MgIrO$_3$. Though the canted-antiferromagnetic state has not been predicted in the theoretical investigation treating a Kitaev model, in the present stage, there is no powerful proof that the observed canting magnetic structure is a new magnetic phase [41]. Even if up and down tilting spins are not antiparallel owing to the presence of space inversion symmetry breaking in IrO$_6$ octahedra, total spin moment are perfectly canceled without weak ferromagnetic moment in the Neel, zigzag and stripe magnetic structures which are predicted in the Kitaev-Heisenberg honeycomb model [38-40]. Thus, the spin tilting cannot generate a weak ferromagnetic moment in this situation. On the other hand, the Dzyaloshinskii-Moriya interaction can make canting of spins, resulting in an appearance of weak ferromagnetic moment. In an ilmenite structure, there is a $D$ vector not on the nearest neighbor interaction ($J_1$) but on the next nearest neighbor one ($J_2$) because of the presence/absence of inversion symmetry. This fact indicates that the Dzyaloshinskii-Moriya interaction in $J_2$ in MgIrO$_3$ would be larger than that in ZnIrO$_3$ since the magnitude of $D$ is in proportion to the magnitude of $J_2$. Indeed, a frustration index $f = |\theta_w|/T_N$ in MgIrO$_3$ is larger than that in ZnIrO$_3$, indicating that $J_2$ in MgIrO$_3$ is effective. In ZnIrO$_3$ and MgIrO$_3$, Ir$^{4+}$ spins interact via next-nearest-neighbor superexchange interactions through Ir$^{4+}$-O$^{2-}$-O$^{2-}$-Ir$^{4+}$ and Ir$^{4+}$-O$^{2-}$-$M^{2+}$-O$^{2-}$-Ir$^{4+}$ ($M$ = Zn, Mg). The magnitude of superexchange interaction through Ir$^{4+}$-O$^{2-}$-O$^{2-}$-Ir$^{4+}$ in both compounds should be comparable, which could not be the origin of different $J_2$. Thus, it can be considered that the origin of difference in the exchange interaction between MgIrO$_3$ and ZnIrO$_3$ is the difference of the Ir$^{4+}$-O$^{2-}$-$M^{2+}$-O$^{2-}$-Ir$^{4+}$ passes with the filled outermost 3$d$ and 2$p$ orbitals, respectively in Zn and Mg ions at the midpoint of the superexchange transferring. From this fact, $J_2$ in ZnIrO$_3$ could be much smaller than that in MgIrO$_3$, which results in that spins in MgIrO$_3$ is more frustrated than that in ZnIrO$_3$. The frustration effect could be also the origin to make noncollinear spin structure with a weak ferromagnetic moment.

Finally, we discuss the different magnetism between the present and the other Kitaev compounds. MgIrO$_3$ and ZnIrO$_3$ show magnetic orderings at a relative higher temperature than Li$_2$IrO$_3$ and Na$_2$IrO$_3$. In a pure Kitaev model, spins show no magnetic ordering [2]. On the other hand, a presence of Heisenberg interaction, which is enhanced by a deviation of the bond angle of Ir-O-Ir from the ideal 90°, gives rise to a magnetic ordering at a finite temperature [41]. In this Kitaev-Heisenberg model, it is expected that a frustration index $f = |\theta_W|/T_N$ becomes larger with being close to the Kitaev limit. For comparison, the bond angle of Ir-O-Ir ($\varphi$) and a frustration index ($f$) in some Kitaev honeycomb iridates are listed in TABLE III. There is no systematic relationship between $\varphi$ and $f$. Both curves of $M/H$-$T$ in $A_2$IrO$_3$ and RuCl$_3$ exhibit a broad peak above $T_N$, which is characteristic in a low dimensional system, while no broad peak appears in $M/H$-$T$ curves in MgIrO$_3$ and ZnIrO$_3$. That is, the magnetic model of $A_2$IrO$_3$ and RuCl$_3$ can be regarded as a quasi-two-dimensional honeycomb lattice, while those in MgIrO$_3$ and ZnIrO$_3$ three-dimensionally layered honeycomb one. Thus, in the case of $A_2$IrO$_3$ and RuCl$_3$, the low dimensionality would contribute the suppression of the magnetic ordering in addition of the frustration effect. This is contrast to the cases in MgIrO$_3$ and ZnIrO$_3$, in which the interlayer cations possibly mediate the interlayer magnetic interactions. However, despite a three-dimensionality, both MgIrO$_3$ and ZnIrO$_3$ shows a tilting magnetic structure similar to the other Kitaev compounds. This fact suggests that the universality in the Kitaev magnetism keeps against an additional non-Kitaev interaction.

**Summary**

We have succeeded in synthesizing the novel honeycomb lattice iridates ZnIrO$_3$ and MgIrO$_3$ via a metathetical decomposition. These compounds have $J_{eff}$ = 1/2 systems of the honeycomb lattice with dominant antiferromagnetic interactions. ZnIrO$_3$ and MgIrO$_3$ show antiferromagnetic ordering probed by the magnetic and thermodynamic measurements, and in the case of MgIrO$_3$, the weak ferromagnetic behavior is observed. Measurement with using the magnetic oriented sample revealed an XY magnetic anisotropy and the magnetic ordering with spin-tilting which is probably stabilized by an effective Kitaev interaction both in

ZnIrO$_3$ and MgIrO$_3$. Moreover, a weak ferromagnetic behavior is observed only in MgIrO$_3$, whose origin could be the Dzyaloshinskii-Moriya interaction and the spin frustration on the honeycomb lattice. We believe that an exotic spin-orbit physics is realized in the newly discovered Ir-honeycomb antiferromagnetic system.

**Acknowledge**

This work is supported by Grant-in-Aids for Scientific Research from MEXT of Japan [Grants Nos. 26410089, 16H04131, and 16J04048].

**Reference**

[1] W. Witczak-Krempa, G. Chen, Y. B. Kim, and L. Balents, Annu. Rev. Condens. Matter Phys. **5**, 57 (2014).
[2] A. Kitaev, Ann. Phys. **321**, 2 (2006).
[3] X.Wan, A. M. Turner, A. Vishwanath, and S. Y. Savrasov, Phys. Rev. B **83**, 205101 (2011).
[4] Y. Okamoto, M. Nohara, H. Aruga-Katori, and H. Takagi, Phys. Rev. Lett. **99**, 137207 (2007).
[5] M. J. Lawler, A. Paramekanti, Y. B. Kim, and L. Balents, Phys. Rev. Lett. **101**, 197202 (2008).
[6] A. Shitade, H. Katsura, J. Kuneŝ, X.-L. Qi, S.-C. Zhang, and N. Nagaosa, Phys. Rev. Lett. **102**, 256403 (2009).
[7] B. J. Kim, H. Ohsumi, T. Komesu, S. Sakai, T. Morita, H. Takagi, and T. Arima, Science **323**, 1329 (2009).
[8] B.J. Kim, Hosub Jin, S. J. Moon, J.-Y. Kim, B.-G. Park, C.S. Leem, Jaejun Yu, T.W. Noh, C. Kim, S.-J. Oh, J.-H. Park, V. Durairaj, G. Cao, and E. Rotenberg, Phys. Rev. Lett. **101**, 076402 (2008)
[9] G. Jackeli and G. Khaliullin, Phys. Rev. Lett. **102**, 017205 (2009).
[10] W. Witczak-Krempa and Y.-B. Kim, Phys. Rev. B **85**, 045124 (2012).
[11] K. Ohgushi, H. Gotou, T. Yagi, Y. Kiuchi, F. Sakai and Y. Ueda: Phys. Rev. B **74**, 241104 (2006).
[12] K. Ohgushi, J. Yamaura, H. Ohsumi, K. Sugimoto, S. Takeshita, A. Tokuda, H. Takagi, M. Takata, and T. Arima, Phys. Rev. Lett. **110**, 217212 (2013).
[13] Y. Singh, S. Manni, J. Reuther, T. Berlijn, R. Thomale, W. Ku, S. Trebst, and P. Gegenwart, Phys. Rev. Lett. **108**, 127203 (2012).
[14] H. Gretarsson, J. P. Clancy, X. Liu, J. P. Hill, E. Bozin, Y. Singh, S. Manni, P. Gegenwart, J. Kim, A. H. Said, D. Casa, T. Gog, M. H. Upton, H.-S. Kim, J. Yu, V. M. Katukuri, L. Hozoi, J. van den Brink, and Y.-J. Kim, Phys. Rev. Lett. **110**, 076402 (2013).
[15] H. S. Chun, J.-W. Kim, J. Kim, H. Zheng, C. C. Stoumpos, C. D. Malliakas, J. F. Mitchell, K. Mehlawat, Y. Singh, Y. Choi, T. Gog, A. Al-Zein, M. M. Sala, M. Krisch, J. Chaloupka, G. Jackeli, G. Khaliullin, and B. J. Kim, Nature Phys. **11**, 462 (2015).
[16] J. G. Rau, E. K.-H. Lee, and H.-Y. Kee, Annu. Rev. Condens. Matter Phys. 7, 195 (2016).
[17] I. Pollini, Phys. Rev. B **53**, 12769 (1996).
[18] K. W. Plumb, J. P. Clancy, L. J. Sandilands, V. V. Shankar, Y. F. Hu, K. S. Burch, H.-Y. Kee, and Y.-J. Kim, Phys. Rev. B **90**, 041112 (2014).
[19] H.-S. Kim, V. Shankar, A. Catuneanu, and H.-Y. Kee, Phys. Rev. B **91**, 241110 (2015).
[20] L. J. Sandilands, Y. Tian, K. W. Plumb, Y.-J. Kim, and K. S. Burch, Phys. Rev. Lett. **114**, 147201 (2015).
[21] Y. Kubota, H. Tanaka, T. Ono, Y. Narumi, and K. Kindo, Phys. Rev. B **91**, 094422 (2015).
[22] L. J. Sandilands, Y. Tian, A. A. Reijnders, H.-S. Kim, K. W. Plumb, Y.-J. Kim, H.-Y. Kee, and K. S. Burch, Phys. Rev. B **93**, 075144 (2016).

[23] A. Koitzsch, C. Habenicht, E. Müller, M. Knupfer, B. Büchner, H. C. Kandpal, J. van den Brink, D. Nowak, A. Isaeva, and T. Doert, Phys. Rev. Lett. **117**, 126403 (2016).
[24] S. C. Williams, R. D. Johnson, F. Freund, S. Choi, A. Jesche, I. Kimchi, S. Manni, A. Bombardi, P. Manuel, P. Gegenwart, and R. Coldea, Phys. Rev. B **93**, 195158 (2016).
[25] S. K. Choi, R. Coldea, A. N. Kolmogorov, T. Lancaster, I. I. Mazin, S. J. Blundell, P. G. Radaelli, Y. Singh, P. Gegenwart, K. R. Choi, S.-W. Cheong, P. J. Baker, C. Stock, and J. Taylor, Phys. Rev. Lett. **108**, 127204 (2012).
[26] F. Ye, S. Chi, H. B. Cao, B. C. Chakoumakos, J. A. Fernandez-Baca, R. Custelcean, T. F. Qi, O. B. Korneta, and G. Cao, Phys. Rev. B **85**, 180403(R) (2012).
[27] J. A. Sears, M. Songvilay, K. W. Plumb, J. P. Clancy, Y. Qiu, Y. Zhao, D. Parshall, and Y.-J. Kim, Phys. Rev. B **91**, 144420 (2015).
[28] S. Hwan Chun, J.-W. Kim, J. Kim, H. Zheng, C. C. Stoumpos, C. D. Malliakas, J. F. Mitchell, K. Mehlawat, Y. Singh, Y. Choi, T. Gog, A. Al-Zein, M. M. Sala, M. Krisch, J. Chaloupka, G. Jackeli, G. Khaliullin, and B. J. Kim, Nat. Phys. **11**, 462 (2015).
[29] H. B. Cao, A. Banerjee, J.-Q. Yan, C. A. Bridges, M. D. Lumsden, D. G. Mandrus, D. A. Tennant, B. C. Chakoumakos, and S. E. Nagler, Phys. Rev. B **93**, 134423 (2016).
[30] A. Banerjee, C.A. Bridges, J.-Q. Yan, A.A. Aczel, L. Li, M.B. Stone, G.E. Granroth, M.D. Lumsden, Y. Yiu, J. Knolle, S. Bhattacharjee, D.L. Kovrizhin, R. Moessner, D. A. Tennant, D.G. Mandrus, and S.E. Nagler, Nat. Mater. **15**, 733 (2016).
[31] S.-H. Baek, S.-H. Do, K.-Y. Choi, Y.S. Kwon, A. U.B. Wolter, S. Nishimoto, J. van den Brink, and B. Büchner, Phys. Rev. Lett. **119**, 037201 (2017).
[32] K. Kitagawa, T. Takayama, Y. Matsumoto, A. Kato, R. Takano, Y. Kishimoto, S. Bette, R. Dinnebier, G. Jackeli, H. Takagi, Nature(London), **554**, 341 (2018).
[33] S. Bette, T. Takayama, K. Kitagawa, R. Takano, H. Takagi, and R. E. Dinnebier, Dalton Trans. **46**, 15216 (2017).
[34] F. Izumi, K. Momma, Solid State Phenom. **130**, 15-20 (2007).
[35] P. Scherrer, N. G. W. Gottingen, Math-Pys. Kl. **2**, 96−100 (1918).
[36] N. E. Brese, M. O'Keeffe, Acta Crystallogr., Sect. B: Struct. Sci **B47**, 192 (1991)
[37] G. Cao, T. Qi, L. Li, J. Terzic, V. Cao, S. Yuan, M. Tovar, G. Murthy, and R. Kaul, Phys. Rev. B **88**, 220414 (2013).
[38] H. C. Jiang, Z. C. Gu, X. L. Qi, and S. Trebst, Phys. Rev. B **83**, 245104 (2011).
[39] J.G. Rau, E. K. H. Lee, and H. Y. Kee, Phys. Rev. Lett. **112**, 077204 (2014).
[40] I. Kimchi and Y. Z. You, Phys. Rev. B **84**, 180407(R) (2011).
[41] J. Chaloupka, G. Jackeli, and G. Khaliullin, Phys. Rev. Lett. 105, 027204 (2010).

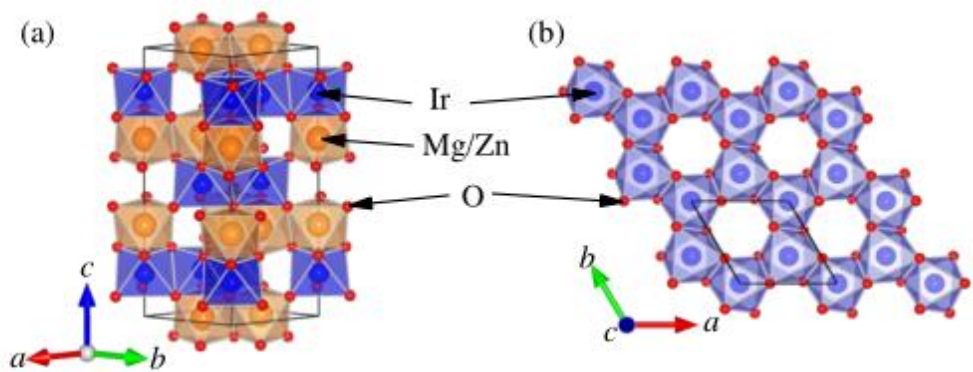

FIG. 1. (Color online) (a) Crystal structure of ZnIrO$_3$ and MgIrO$_3$. (b) IrO$_3$ layer viewed along the $c$-axis.

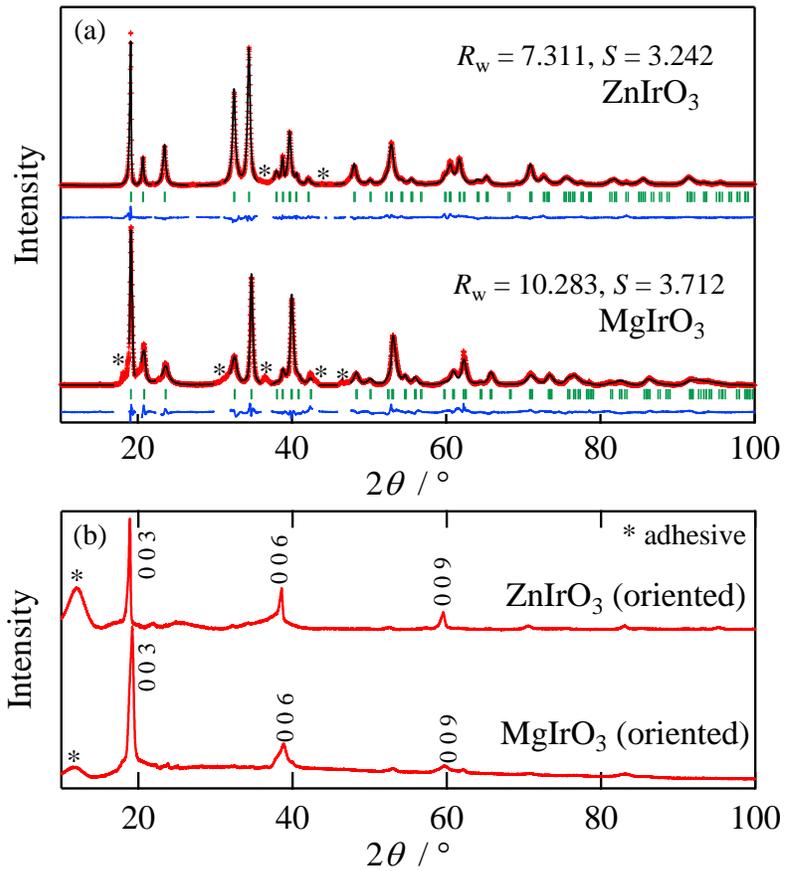

FIG. 2. (Color online) (a) XRD patterns of powder samples in $ZnIrO_3$ and $MgIrO_3$ measured at room temperature. Vertical bars indicate the positions of Bragg reflections. Asterisks indicate unknown impurities. (b) XRD patterns of oriented powder samples in $ZnIrO_3$ and $MgIrO_3$. Numbers above the simulated profile designate indices of each diffraction. Asterisks indicate the peak derived from the adhesive glue.

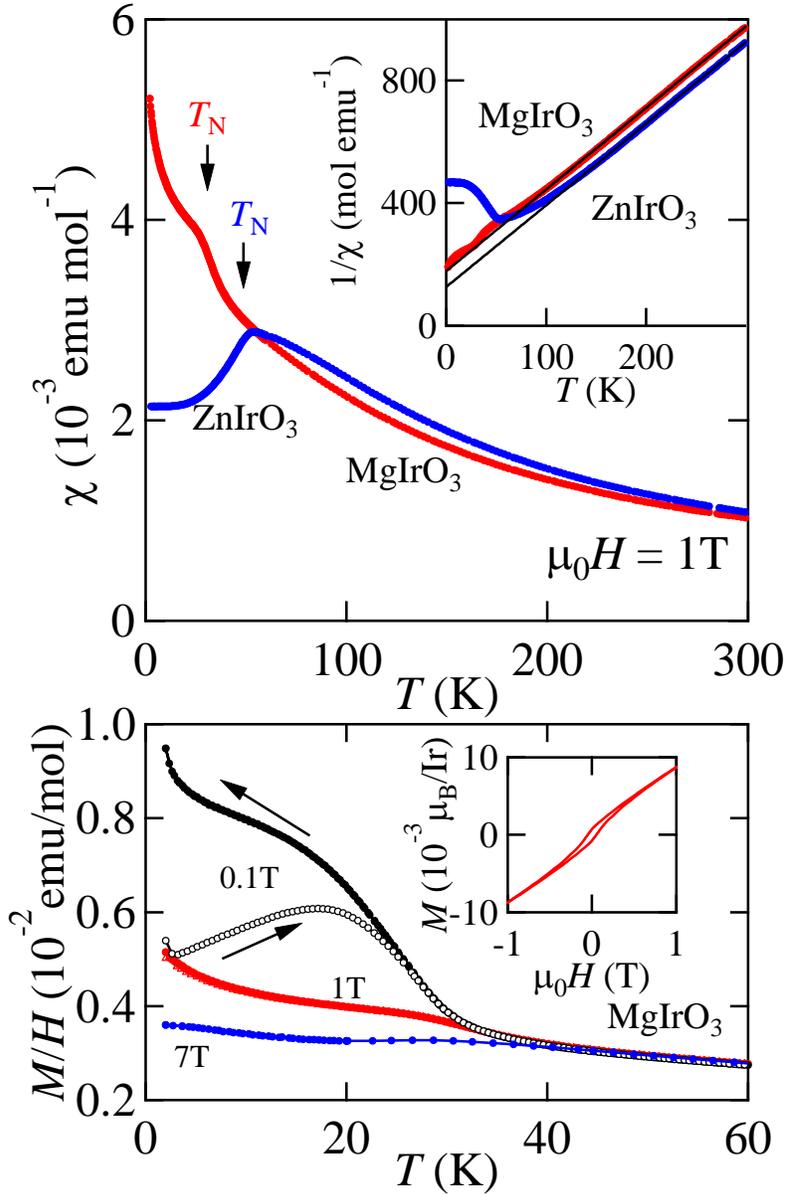

FIG. 3. (Color online) (a) Temperature dependence of magnetic susceptibility χ of ZnIrO$_3$ and MgIrO$_3$. The inset shows the inversed magnetic susceptibility 1/χ. (b) The temperature dependence of the magnetization divided by the external field $M/H$ in MgIrO$_3$ measured at several magnetic fields. The inset shows the hysteresis loop of the magnetization $M$.

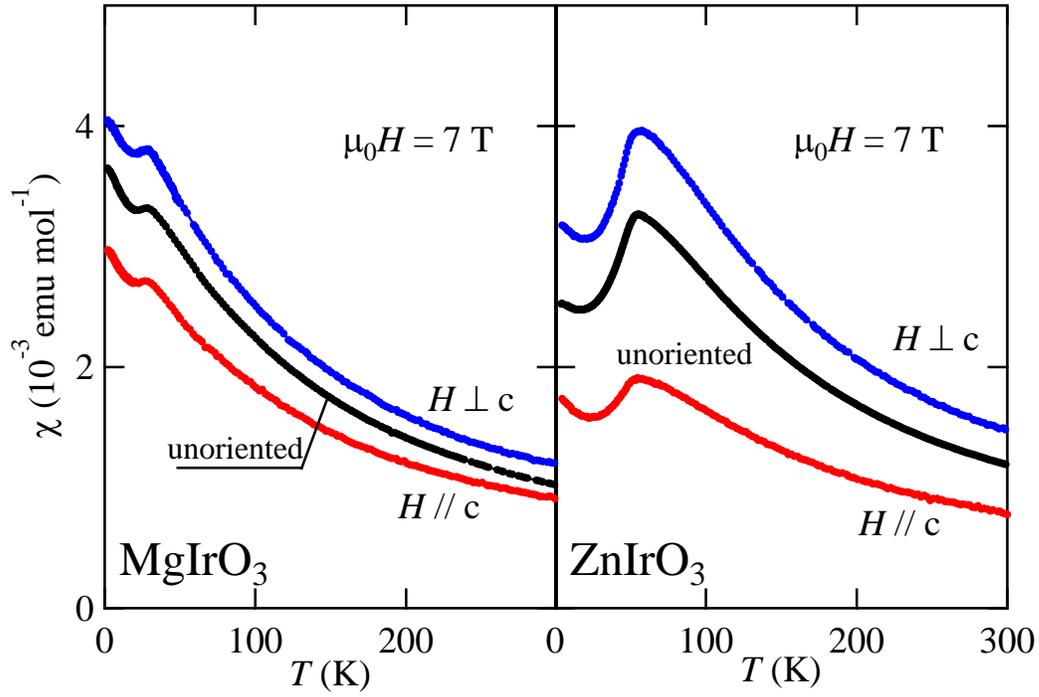

FIG. 4. (Color online) The temperature dependence of magnetic susceptibility in ZnIrO$_3$ and MgIrO$_3$ measured using unoriented and oriented powder samples.

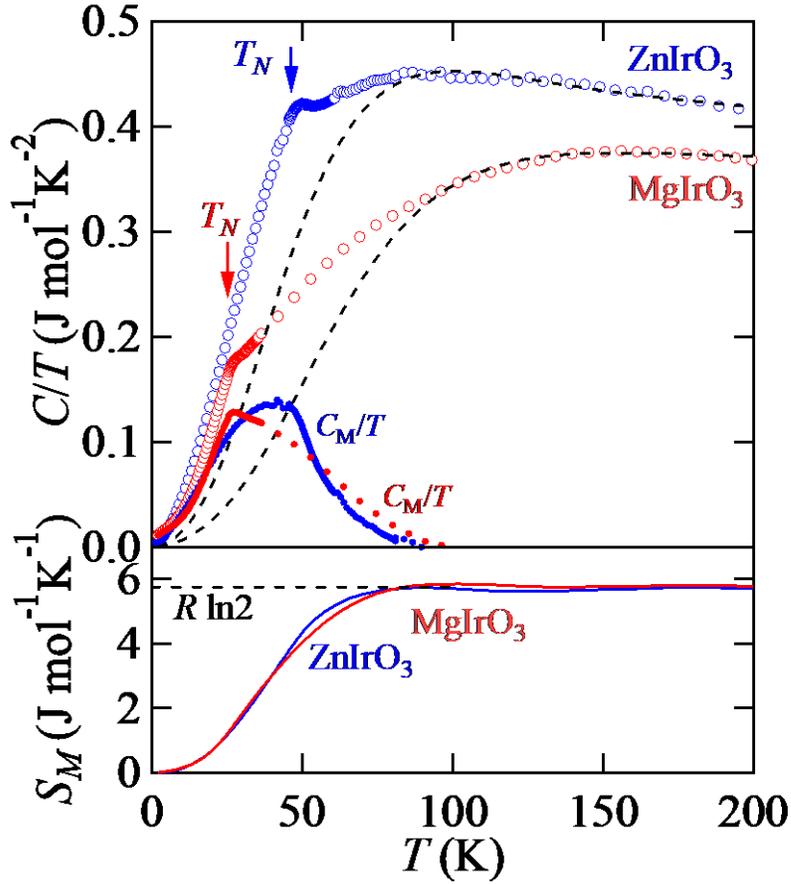

FIG. 5. (Color online) Upper panel shows the temperature dependence of the heat capacity divided by temperature $C/T$ and the magnetic heat capacity $C_M/T$ after subtracted the lattice contribution in ZnIrO$_3$.and MgIrO$_3$. The dashed lines represent the lattice contributions estimated by fitting the data above 100 K, as described in the text. The downer panel shows the temperature dependence of the magnetic entropy $S_M$ obtained by integrating the $C_M/T$ data as a function of temperature.

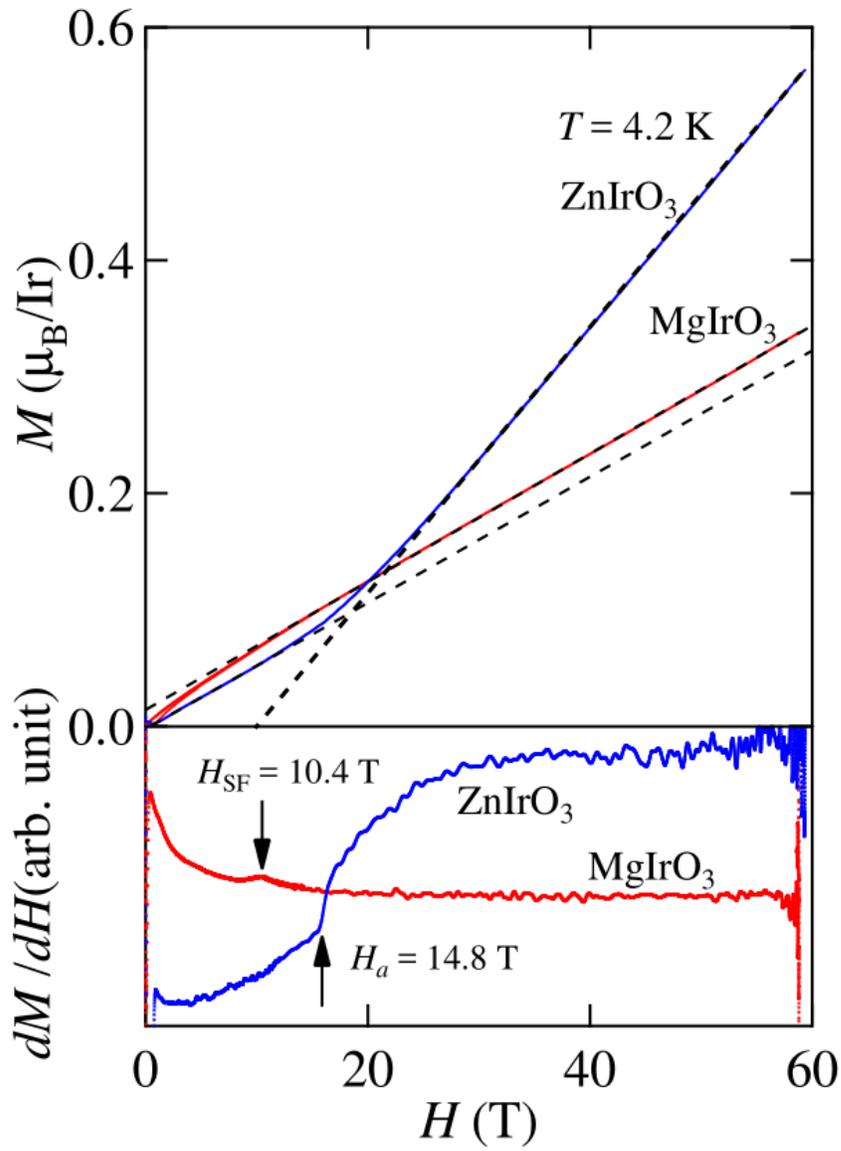

FIG. 6. (Color online) Magnetization ($M$) and differential susceptibility ($dM/dH$) curves of $ZnIrO_3$ and $MgIrO_3$.

TABLE I. Crystallographic parameters for $ZnIrO_3$ and $MgIrO_3$ (both $R$-3) determined using powder x-ray diffraction.
The obtained lattice parameters are $a = 5.1992(3)$ and $5.1584(5)$ Å and $c = 13.8903(6)$ and $13.919(1)$ Å, respectively, for $ZnIrO_3$ and $MgIrO_3$. $B$ is the thermal displacement parameter.

|  | site | $x$ | $y$ | $z$ | $B$ (Å) |
|---|---|---|---|---|---|
| $ZnIrO_3$ |  |  |  |  |  |
| Zn1 | 6c | 0 | 0 | 0.3715(1) | 0.34 |
| Ir1 | 6c | 0 | 0 | 0.1600(1) | 0.38 |
| O1 | 18f | 0.3355(11) | -0.0077(10) | 0.0938(3) | 0.27 |
| $MgIrO_3$ |  |  |  |  |  |
| Mg1 | 6c | 0 | 0 | 0.3499(6) | 0.18 |
| Ir1 | 6c | 0 | 0 | 0.1587(1) | 0.34 |
| O1 | 18f | 0.3226(17) | 0.0065(13) | 0.0848(5) | 0.20 |

TABLE II. Experimentally obtained Curie-Weiss Parameters and anisotropic $g$-factors in ZnIrO$_3$ and MgIrO$_3$.

| formula | $p_{\text{eff}}$ | $\theta_W$ (K) | $g_\perp$ | $g_\parallel$ | $g_\perp/g_\parallel$ |
|---|---|---|---|---|---|
| ZnIrO$_3$ | 1.733(7) | -47.5(6) | 2.194(2) | 1.494(4) | 1.468 |
| MgIrO$_3$ | 1.734(6) | -67.1(5) | 2.070(3) | 1.907(1) | 1.085 |

TABLE III. The bond angle $\varphi$ of Ir-O-Ir and a frustration index $f$ in the Kitaev honeycomb iridates.

| sample | $\varphi$ (°) | $f = |\theta_W|/T_N$ | reference |
|---|---|---|---|
| ZnIrO$_3$ | 95.7 | 1.02 | this work |
| MgIrO$_3$ | 94.0 | 2.11 | this work |
| Li$_2$IrO$_3$ | 94.7, 95.3 | 4.07 | [13] |
| Na$_2$IrO$_3$ | 98.0, 99.4 | 7.33 | [13] |